\documentclass[aps,pre,twocolumn,amsmath,amssymb,showpacs]{revtex4}
\usepackage{bm}
\usepackage{graphicx}

\newcommand{\be}{\begin{eqnarray}}
\newcommand{\ee}{\end{eqnarray}}

\newcommand{\ba}{\begin{array}}
\newcommand{\ea}{\end{array}}

\newcommand{\ket}[1]{|#1\rangle}

\begin{document}

\title{How not to discard half of the cases in QKD}
\author{Marcin Paw{\l}owski}
\address{ Institute of Theoretical Physics and Astrophysics, Uniwersytet Gda\'{n}ski, PL-80-952, Gda\'{n}sk
\\
Katedra Fizyki Teoretycznej i Informatyki Kwantowej, Politechnika Gda\'{n}ska, PL-80-952, Gda\'{n}sk}

\date{13 August 2007}

\begin{abstract}
All known QKD protocols require the parties to discard the results when they have chosen different bases.
In this paper we show that it is not necessary. We give examples of QKD protocols that are as safe as standard ones
but do not involve the discarding of the results when the bases are different or even that do not require the announcement of
the bases. This leads to greater communication channel capacities and render some eavesdropping strategies useless
but the most important thing is that they provide
us better insight into the general structures that underlie the quantum cryptography and help to
establish the boundaries of what is possible and what is not.
\end{abstract}


\maketitle

All known quantum key distribution (QKD) protocols, whether entanglement based like Ekert's \cite{E91} or not like BB84 \cite{BB84}, share
one common trait. The communicating parties choose some measurement bases and later announce them to find whether or not did they manage
to establish one bit of secret key. In most of the protocols they can do this in one half of the cases. In the rest of the protocols this
ratio is similar or less \cite{Gisin}. The only exception is the efficient protocol described in \cite{Lo}.
Though the parties discard
the cases when they have chosen the different bases this happens very rarely
due to the fact that the bases are chosen with different
probabilities. This leads to the similar advantages that protocols described in this paper
have when it comes to communication channel capacities but, since
the bases are still announced by both parties,
does not provide the protocol presented in \cite{Lo} with similar advantages when it comes to
security issues.
 Before we present the protocols that do not require discarding of the cases when the chosen bases are different, let us take
a look at the QKD from the probability distribution point of view.

All QKD protocols can be viewed as follows. There is a joint probability distribution $p(a,b,A,B)$. Each of the two parties \cite{3P} has
knowledge of the two of the variables (Alice knows her choice of the basis $a$ and the measurement outcome or the value of the bit coded $A$ and
Bob knows $b$ and $B$). The protocols are constructed in such a way that knowledge of two variables gives nothing, while the knowledge of three
gives the fourth, but not in all the cases. There are combinations of variables values for which the fourth one is automatically known and
there are combinations that give no information about it. Later the parties publicly announce one of their variables $a$ and $b$,
so now they both have the necessary knowledge of three variables, while anyone listening to the public channel has only two.
They also know whether this combination of variables leads to the fourth one (this happens if $a=b$). If the bases are randomly chosen, then
from the probability distribution point of view each variable has the same status, but in all existing protocols variables denoted by lower case letters
(bases) are preferred. We conjecture (and prove) that giving equal rights to all variables can be beneficial
to quantum cryptography.

First let our parties reveal $A$ and $B$ (that is the outcomes) instead of $a$ and $b$ (that is the bases). Let them do that in the standard BB84 protocol.
The outcomes will be the same in 75\% of the cases (always if they have chosen the same bases and in one half of the cases when they have chosen different
ones), these runs of the protocol will be discarded. If the outcomes are different then they are sure that they have chosen the different bases
and they can use that to generate the bit of the key. Listening to the public channel gives the third party nothing, since being sure that the
bases are different is not enough, knowledge of $a$ or $b$ is also necessary. This naive example gives us no benefit when it comes to channel capacities \cite{2}.
On the contrary instead of discarding 50\% of the cases we discard 75\%, but it provides us with some important insight. We see that bases do not have to
be necessarily announced. And since they do not, there may be no need to discard the cases when they are different.

Before presenting the QKD protocol that does not require the parties to discard almost anything
let us take a look at a Bell inequality that will provide safety criterion.
Let us consider the simplified version of CGLMP inequality \cite{CGLMP} given by Zohren and Gill \cite{ZG}
\be \label{ZG} \nonumber
A_d(\psi)= & P(A_2<B_2)+P(B_2<A_1) &
\\
&+P(A_1<B_1)+P(B_1\leq A_2)&\geq 1
\ee
Here $X_i$ denotes the measurement outcome of party $X$ in the basis $i$. The outcomes can take the value 0,1,..,$d-1$.
In \cite{ZG} it is shown that for quantum probabilities this inequality is violated and as the dimensionality of the systems $d$ rises
there are states for which $A_d$ approaches zero. Surprisingly, these optimal states are not maximally entangled. For optimal states and
huge $d$'s this implies that one can find measurement bases for which
\be
P(A_2 \geq B_2) \approx 1, \quad P(A_1 \geq B_1) \approx 1
\\
P(B_2 \geq B_1) \approx 1, \quad P(B_1 > A_2) \approx 1
\ee
Which means that the following is almost always true
\be \label{3}
B_2 \geq A_1 \geq B_1
\\ \label{4}
B_1 > A_2 \geq B_2
\ee
If Alice and Bob announce the measurement outcomes and discard the cases when they both announce the same number (which will be a very rare event),
they know their choices of bases. Explicitly, if Alice has the greater number they both know that they have chosen the base with the same index
(note that $A_1$ is a different base than $B_1$), if Bob has the greater number they have chosen the bases with different indexes. It is clear that
this phenomenon can be exploited to create quantum key distribution protocol.

One of the possibilities for the QKD is:

One subsystem of the state $\ket{\psi}=\sum_{i=0}^{d-1}\lambda_i\ket{ii}$ (which $\lambda$'s are chosen in such a way that
$\ket{\psi}$ maximally violates the inequality (\ref{ZG}))
is being send to each party, which measures it in one of the bases composed of the vectors
\begin{eqnarray}
\ket{i}_{A,a} & = &\frac{1}{\sqrt{d}} \sum_{k=0}^{d-1} \exp\left(\mathbf{i}\frac{2\pi}{d}k (i+\alpha_a) \right) \ket{k}_A \label{eq:bestmeasA},\\
\ket{j}_{B,b} & = & \frac{1}{\sqrt{d}} \sum_{l=0}^{d-1} \exp\left(\mathbf{i}\frac{2\pi}{d}l (-j+\beta_b) \right) \ket{l}_B \label{eq:bestmeasB},
\end{eqnarray}
$a,b=1,2$ is the index of the chosen basis $\alpha_1=0$, $\alpha_2=\frac{1}{2}$, $\beta_1=\frac{1}{4}$ and $\beta_2=-\frac{1}{4}$.
They announce their results but keep the choices of the bases secret. The only (very rare) case when they discard the run of the protocol is
when they both announce the same outcome. The choice of basis of one previously established party (say Alice)
is going to be the key if the eavesdropping check finds nothing. For this check the parties randomly choose some runs of the protocol, announce
their choices of the bases and compute the value $A_d(\psi)$. Since $\ket{\psi}$ is the optimal state and the bases are complementary, any
measurement or substitution of another state necessarily leads to Alice and Bob receiving altered (not optimal) sate for which the value of $A_d(\psi)$
is greater. If this value differs too much from the expected the presence of the eavesdropper is assumed and the protocol is aborted.

This protocol has several advantages. If $d$ is big enough, almost certainly no cases will be discarded and this whole step can be omitted.
In this case every run of the experiment generates one bit of the key (if we do not count some runs needed for the eavesdropping test).
Furthermore if the Alice's choice of the basis is going to be the key than it is possible for Bob to send only one bit to Alice on the public channel
while the rest of the classical communication is done one-way (Alice to Bob), which can be beneficial in some cases. To do so only Alice
announces the outcomes (which further complicates the job of eavesdropper) of her measurements and Bob always can find what her base was. Next Alice
chooses randomly some of the cases and sends both variables to Bob, who computes $A_d(\psi)$ and sends back his only bit, which carries the information
whether the eavesdropper was present or not.

Unfortunately this protocol has also one serious disadvantage. It is highly impractical, since it works fine only for extremely huge $d$'s. For
small $d$'s (\ref{3}) and (\ref{4}) are not so often true and when they are the outcomes are usually equal. The situation gets better as $A_d(\psi)$
goes to zero but it falls off slower than logarithmically with the dimension \cite{ZG}. The dimensions when it starts to be better than standard ones
are of the order $10^4$, which makes its physical realization far beyond the scope of the current technology. Nevertheless, it proves that it is possible
to construct a protocol that does not require any run to be discarded.

The question arises. It is possible to construct a protocol that does not require any run to be discarded.. The role of bases and the measurement
outcomes can be swapped. Is it possible to construct a QKD protocol that involves announcement of the bases but does not require us to discard
anything? The answer again is: yes.

Consider a following protocol:
\

Alice encodes a random dit in one of the two mutually unbiased bases, given by vectors (\ref{eq:bestmeasA}). That is, if she wants to encode a dit
with value $i=0,1,..,d-1$ in the base $a=1,2$, then she prepares the system in the state $\ket{i}_{A,a}$.
She sends it to Bob. For each system received Bob randomly chooses one of the two modes of operation. The first mode will be
used for the key generation, while the second one will be used for the eavesdropping detection.
\begin{itemize}
\item
If Bob chooses
the key mode of operation, he randomly chooses one of the bases given by vectors (\ref{eq:bestmeasB}) and makes his measurement.
Note that in this mode Bob always measures system in a different base than it was prepared. Nevertheless, we will show that Bob is
able to decode the dit of the key with the probability high enough to make this protocol effective.
\item
In the security mode he randomly chooses one of the bases that Alice could have encoded her dit in (\ref{eq:bestmeasA}) and measures the system.
\end{itemize}
After all the systems are sent and received Alice announces her choice of the bases and Bob announces his choice of the modes.
All the cases when Bob had chosen the security mode are used to check for eavesdropping
in a standard manner. That is, both parties announce their results and bases. If they chose the same bases they should get the same
outcomes. If this is not the case, they can assume that they had been eavesdropped. This mode provides the security against eavesdropping which works exactly
as in the protocol presented in \cite{Multilevel}.

How does Bob decode the dit when he has measured the system in a different base that it has been prepared? It happens that knowing both of the bases
and his result Bob knows that one of the possible letters of the alphabet is much more probable than the others. The probability that Bob gets result
$j$, when he had chosen base $b$ and Alice prepared the state $\ket{i}_{A,a}$ is
\be \label{prob}
p(a,b,i,j)=\frac{1}{d\Big[(1-\cos \left(\frac{2\pi}{d}(i+j+\alpha_a-\beta_b) \right)\Big]}
\ee

If $a+b<4$ then the most probable outcomes are those satisfying $i+j=0 \mod d$. If $a=b=2$ then they satisfy $i+j=-1 \mod d$.

Note that the structure of (\ref{prob}) assures that we have only $d$ possible values of $p(a,b,i,j)$ and
for each $j$
the set of possible values is the same regardless
of the choice of $a,b$ and $i$.

We end up with situation in which no cases are discarded but the structure of the protocol itself introduces some errors in the key. To answer whether
or not this protocol allows for the faster key generation we need to look at the channel capacities.

Classically noiseless channel with $d$-dimensional input and output alphabets has the capacity $\log d$. When it comes to QKD, even with
perfect quantum channels this value drops to $\frac{1}{2}\log d$. It is due to the fact that the channels used in cryptography can be viewed
as erasure channels, that is channels that do not change the input with the probability $1-p$ and erase it with the probability $p$. In standard QKD
$p=50\%$ which corresponds to one half of the cases when Alice and Bob have chosen different bases and had to discard (erase) their results.
\

For the sake of clarity let us look at BB84 protocol from the point of view of information channels. The bit encoded by Alice in
some quantum two-level system is send to Bob by a channel $\mathcal{N}_s=\mathcal{N}_2\circ\mathcal{N}_1$. $\mathcal{N}_1$ is a quantum channel
which we will assume to be perfect and not deal with. $\mathcal{N}_2$ is a classical erasure channel with the capacity $\frac{1}{2}$. Since
$\mathcal{N}_1$ is noiseless, $\mathcal{N}_s$ also has the capacity of $\frac{1}{2}$. For qdits in general this becomes the mentioned $\frac{1}{2}\log d$.

\

If
we again assume that the quantum part of the channel is perfect then the capacity of the channel $\mathcal{N}_2$ which is the same as the capacity of
the whole channel $\mathcal{N}$ can be easily computed (see footnote \cite{comp})
\be \label{cap}
C(\mathcal{N})=2 \log d + \sum_{a,b=1}^2\sum_{i,j=0}^{d-1}\frac{p(a,b,i,j)}{4d} \log \frac{p(a,b,i,j)}{4d}
\ee

This capacity is greater than $C(\mathcal{N}_s)=\frac{1}{2}\log d$ for $d>3$ as  Fig. 1 shows.

\begin{figure}[!h]
\includegraphics[scale=0.3]{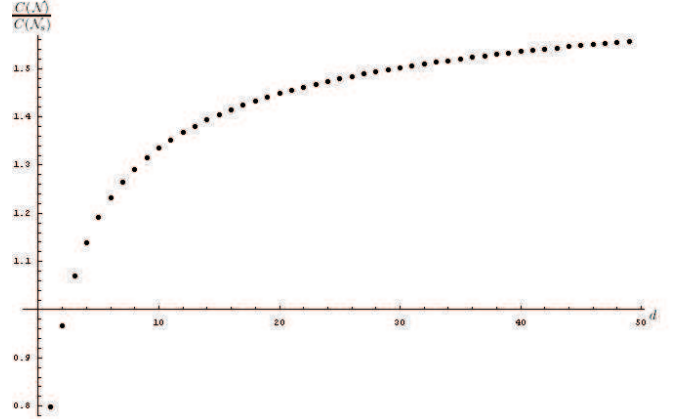}
\caption{The ratio of channel capacities for standard QKD and the "practical" protocol presented in this paper as a function of the Hilbert space dimension}
\end{figure}

 Introducing some noise in the part
of the channel $\mathcal{N}_2$ does not change anything in the analysis, since it affects the standard protocols and this one in exactly the same way.

Note that this protocol can be slightly modified to have another interesting property. If Bob in the key generation mode always chooses
the base $b=1$ he knows that regardless of Alice's choice $a+b<4$. Then he is sure that Alice's and his most probable outcomes satisfy
$i+j=0~\mod~d$ so there is no need for Alice to announce her basis. We end up with a protocol in which nothing is publicly announced! Unfortunately
this modification allows Eve to introduce less errors by eavesdropping \cite{mod}, but is true only when she is using certain strategies. Other
strategies that relay on the knowledge of the announced variables are not possible at all.

We have presented an approach to quantum cryptography, in which probability distributions are considered and each variable
(setting or outcome) is treated in the same fashion. This allows for more flexibility in protocol design which in turn leads to protocols
with higher capacities and other advantages.
For example, note that in both protocols only one variable (or even none in the modified practical one)
is announced publicly, which can make the eavesdropper work significantly harder (the eavesdropping strategies that require the knowledge of the publicly
announced data are simply not possible).
QKDs presented in this paper are by no means
the end to the possibilities that arise before quantum cryptography if we choose to look at it from a different angle.

The author would like to thank S. Zohren for sharing his numerical data, M. \.{Z}ukowski for helping to make this paper more clear and H.-K. Lo
for bringing paper \cite{Lo} to authors attention.
 This work is part of EU 6FP programme QAP contract no. 015848.


\begin{thebibliography}{9}
\bibitem{E91}A. K. Ekert, Phys. Rev. Lett. {\bf 67}, 661 (1991)
\bibitem{BB84}C. H. Bennett and G. Brassard, Proc. IEEE Int. Conf. on Computers, Systems, and Signal Processing, Bangalore, India (New York, IEEE, 1984)
\bibitem{Gisin} N. Gisin, G. Ribordy, W. Tittel, H. Zbinden, Rev. Mod. Phys. {\bf 74}, 145 (2002)
\bibitem{Lo} H.-K. Lo, H.F. Chau, M. Ardehali, J. of Cryptology, {\bf 18}(2), 133-165 (2005)
\bibitem{3P} The third party, or the eavesdropper is not considered here, since exactly the same safety measures as in standard QKD protcols are
employed in each of the protocols presented in this paper. Therefore we can assume that if this party actively existed it had been discovered and
the protocol aborted.
\bibitem{2} The check for eavesdropping is similiar to standard BB84. For some randomly chosen runs of protocol bases are revealed and if Eve had
tampered with trensfered qubits, then sometimes though the results are different the bases are the same, so Eve gets spotted.
\bibitem{CGLMP} D. Collins, N. Gisin, N. Linden, S. Massar, S. Popescu, Phys. Rev. Lett. {\bf 88}(4),040404 (2002)
\bibitem{ZG} S. Zohren, R.D. Gill,  arXiv:quant-ph/0612020v2
\bibitem{Multilevel} M. Bourennane, A. Karlsson, G. Bjork, N. Gisin, N. Cerf, J. Phys. A {\bf 35}, 10065 (2002)
\bibitem{comp} The capacity of the communication channel $\mathcal{N}$ is given by
\be \label{op1}
C(\mathcal{N})=\max_{p(x)}H(X:Y)
\ee
where $p(x)$ are input distributions for $X$ and $Y$ is the corresponding induced random variable at the output channel. For QKD the only appropriate
input distribution is the uniform one $p(x)=\frac{1}{d}$ since we want to get a totaly random key. So (\ref{op1}) becomes
$C(\mathcal{N})=H(X)+H(Y)-H(X,Y)$. From this we arrive at (\ref{cap}).
\bibitem{mod} Consider the following strategy for Eve. She intercepts the system and measures it in the basis (\ref{eq:bestmeasB}) with $b=1$. After
the measurement she resends the system to Bob in the state that the system collapsed to. If this run of the protocol is used for the security check
then Eve's actions will be detected with the same probability for both modified and unmodified versions of the protocol. But in the modified verison
Eve gets the same outcome as Bob with probablity 100\%. In the unmodified version this probablity is only $\frac{d+1}{2d}$, so Eve needs more attacks
to gain the same amount of the information about the key.
\end{thebibliography}
\end{document}